\documentclass[preprint]{aastex}
\usepackage{CJK}
\usepackage{amssymb}
\usepackage{amsmath}
\usepackage{graphicx}
\usepackage{graphics}
\usepackage{amsfonts, color, hyperref, epsfig}
\newcommand{\kms}{{km~s$^{-1}$}}
\newcommand{\lya}{Ly$\alpha$}
\newcommand{\nv}{N\,{\footnotesize V}}
\newcommand{\civ}{C\,{\footnotesize IV}}
\newcommand{\oiii}{[O\,{\footnotesize III]}}
\newcommand{\oiv}{O\,{\footnotesize IV]}}
\newcommand{\siiv}{Si\,{\footnotesize IV}}
\newcommand{\mgii}{Mg\,{\footnotesize II}}
\newcommand{\feii}{Fe\,{\footnotesize II}}
\newcommand{\ciii}{C\,{\footnotesize III]}}
\newcommand{\siii}{Si\,{\footnotesize III]}}
\newcommand{\aliii}{Al\,{\footnotesize III}}
\newcommand{\ovi}{O\,{\footnotesize VI}}
\newcommand{\hei}{He\,{\footnotesize I}}

\newcommand{\hb}{H$\beta$}
\newcommand{\ha}{H$\alpha$}
\def\J0006{\mbox{SDSS~J0006+1215}}

\shorttitle{SDSS J0006+1215}
\shortauthors{Zhang et al.}

\begin{document}

\title{ Ultraviolet and Optical Emission-line Outflows in the Heavily Obscured  Quasar SDSS J000610.67+121501.2: At the Scale of the Dusty Torus and Beyond}
\author{Shaohua Zhang\altaffilmark{1}, Hongyan Zhou\altaffilmark{1,2}, Xiheng Shi\altaffilmark{1},
Xiang Pan\altaffilmark{1,2}, Ji Wang\altaffilmark{3}, Ning Jiang\altaffilmark{2}, Tuo Ji\altaffilmark{1}, Peng Jiang\altaffilmark{1}, Wenjuan Liu\altaffilmark{4}, Huiyuan Wang\altaffilmark{2}}
\affil{$^1$Polar Research Institute of China, 451 Jinqiao Road, Shanghai, 200136, China; zhangshaohua@pric.org.cn, zhouhongyan@pric.org.cn\\
$^2$Key Laboratory for Research in Galaxies and Cosmology, Department of Astronomy,
    University of Sciences and Technology of China, Chinese Academy of Sciences, Hefei, Anhui 230026, China\\
$^3$California Institute of Technology, 1200 East California Boulevard, Pasadena, CA 91101, USA
$^4$Yunnan Observatories, Chinese Academy of Sciences, Kunming, Yunnan 650011, China}

\begin{abstract}
Broad emission-line outflows of active galactic nuclei (AGNs) have been proposed for many years
but are very difficult to quantitatively study because of the coexistence of the gravitationally-bound and outflow emission.
We present detailed analysis of a heavily reddened  quasar, SDSS J000610.67+121501.2,
whose normal ultraviolet (UV) broad emission lines (BELs) are heavily suppressed by the Dusty Torus as a natural ``Coronagraph'',
thus the blueshifted BELs (BBELs) can be reliably measured.
The physical properties of the emission-line outflows are derived as follows: ionization parameter $U \sim 10^{-0.5}$,
column density $N_{\rm H}\sim 10^{22.0}$ cm$^{-2}$, covering fraction of $\sim 0.1$ and upper limit density of $n_{\rm H}\sim 10^{5.8}$ cm$^{-3}$.
The outflow gases are located at least 41 pc away from the central engine,
which suggests that they have expanded to the scale of the dust torus or beyond.
Besides, \lya\ shows a narrow symmetric component, to our surprise, which is undetected in any other lines.
After inspecting the narrow emission-line region and the starforming region as the origin of the \lya\ narrow line,
we propose the end-result of outflows, diffusing gases in the larger region, acts as the screen of \lya\ photons.
Future high spatial resolution spectrometry and/or spectropolarimetric observation are needed to make a final clarification.

\end{abstract}

\maketitle

\section{Introduction}

The blueshifted emission lines (blueshifted-ELs) and broad absorption lines (BALs), which are the typical presentations of active galactic nuclei (AGNs) outflows, have been widely studied.
They are  potentially one of the most important 
pieces of feedback from AGNs and  may play a crucial role in galaxy formation and evolution,
connecting the central supermassive black holes (SMBHs) to their host galaxies and regulating their co-evolution
(e.g., Granato et al. 2004; Scannapieco \& Oh 2004; Hopkins et al. 2008). Previous works suggested that the AGN outflows
can reach from the  small scale (at or outside of the broad-line region (BELR), e.g., Zhang et al. 2015b; Williams et al. 2016),
to the intermediate scale (the dusty torus, e.g., Leighly et al. 2015; Zhang et al. 2015c; Shi et al. 2016),
the large scale (the inner narrow-line region (NELR), e.g., Komossa et al. 2008; Crenshaw et al. 2010; Ji et al. 2015; Liu et al. 2016),
and even the entire host galaxy of the quasar (e.g., Zakamska et al. 2016).
There is general agreement that there has to be a radial component of motion and some opacity source;
however, questions regarding the origins, whether homologous or not, and the properties themselves of the blueshifted-ELs and BALs remain interesting subjects of debate.
They can also be used as important test beds for models of the BELR, broad absorption line region (BALR), and NELR.

Comparatively, studies on the blueshifted-ELs are more difficult than those on the BALs.
This is because the BALs have high-speed blueshifted (up to a velocity of $v \sim 0.2c$) and broad-width (at least 2000 \kms) troughs
from a wide range of species, such as \mgii, \aliii, and \feii, to \nv, \civ, \siiv, and \ovi\
(e.g., Weymann et al. 1991; Hall et al. 2002; Reichard et al. 2003; Trump et al. 2006; Gibson et al. 2009; Zhang et al. 2010),
and even from the hydrogen Balmer series (e.g., Hall 2007; Ji et al. 2012, 2013; Zhang et al. 2015c) and
metastable \hei\ (e.g., Liu et al. 2015, 2016),
whose obvious features allow the BALs to be easily detected and measured.
Unfortunately, the broad and narrow blueshifted-ELs (BBELs and BNELs) are generally mixed with the normal broad or narrow lines.
The existence of the blueshifted-ELs is expressed by the differences in the line profile (e.g., asymmetry, width, peak and/or velocity centroid)
with respect to the lines emitting from the BELR and NELR (e.g., Gaskell 1982; Wilkes 1984; Crenshaw et al. 1986, 2010;
Marziani et al. 1996, 2013; Richards et al. 2002; Zamanov et al. 2002; Boroson 2005; Snedden \& Gaskell 2007),
and the strength measurement relies on the blueshift and asymmetry index (BAI) for \civ\ (Wang et al. 2011, 2013),
or the distinguishment of the blueshifted wings for \oiii$\lambda$5007 (e.g., Komossa et al. 2008; Zhang et al. 2011; Zakamska et al. 2016)
and for \hb\ and \mgii\ (e.g., Marziani et al. 2013).

Though the fundamental parameters of AGNs that affect the blueshifted-ELs have been found, such as the Eddington ratio (e.g., Komossa et al. 2008; Wang et al. 2011),
the physical properties and  locations of these outflows are hard to be investigate, except for a limited number of sources,
in which the spectra of the blueshifted-ELs from a range of species can be reliably measured.
For example, Leighly et al. (2004a) identified broad, strongly blueshifted high-ionization lines of \civ\ and \nv,
which are dominated by emission from the wind in  
two extreme narrow-line Seyfert 1 galaxies ($FWHM_{\rm MgII}\lesssim$ 1000 \kms): IRAS 13224-3809  and 1H 0707-495.
The photoionization analysis favors a wide range of densities (${\rm log}~n_H\sim7-11\rm~cm^{-3}$)
and a relatively high ionization parameter (${\rm log}~U\sim-1.2~{\rm to}~-0.2$). Moreover,
the blueshifted-EL emission regions and the BELRs do not differ in the location, as $\sim 10^4~r_S$ was inferred for both (Leighly et al. 2014b).
Liu et al. (2016) identified the above three phenomena produced by the AGN outflows in the optical to near-infrared (NIR) spectra
of SDSS J163459.82+204936.0, i.e., a cuspy BNEL component in \hb, \hei$\lambda\lambda$5876,10830,
a BBEL component of \hei$\lambda$10830, the bulk blueshifting of \oiii$\lambda$5007, and BALs in \ion{Na}{1} D and \hei$\lambda\lambda$5876,10830.
The physical parameters determined with $Cloudy$ for absorption-line and emission-line outflows are very close, with
${\rm log}~n_H \sim 4.5-5\rm~cm^{-3}$ and ${\rm log}~U\sim-1.3~{\rm to}~-1.0$, and the distance of the outflow materials are $48-65\rm~pc$ from  the central, exterior of the torus. Of special interest is the similarity of the physical parameters and the locations,
which strongly suggest blueshifted-ELs and BALs should be generated in the common outflowing gas.

In a study of the intermediate-width emission lines (IELs),
Li et al. (2015) used the Dusty Torus as a ``Coronagrap'', which heavily suppress the normal broad emission lines (BELs) in OI 287,
thus, the ultraviolet (UV) IELs can be reliably detected from the quasar emission-line spectra.
This work reminds to us that partially obscured quasars may provide an opportunity to reliably detect the blueshifted-ELs.
If the blueshifted-ELs does exist, they would become prominent in the shorter wavelength range
where the normal BELs are heavily suppressed.
In this work, we report such a infrared-luminous quasar J000610.67+121501.2 (hereafter \J0006),
at $z=2.3174$ with outflows revealed in \civ, \lya, and \nv\ BBELs.
The organization of this paper is as follows. The data we used will be described in Section 2.
We will fit the spectrum and analyze the UV BBELs in Section 3, and discuss the properties
and possible origins of BBEL outflows in Section 4. A summary of our results will be given in Section 5.
Throughout this paper, we adopt the cold dark matter `concordance' cosmology
with H$_{\rm 0}$ = 70 km s$^{-1}$Mpc$^{-1}$, $\Omega_{\rm m} = 0.3$, and $\Omega_{\Lambda} = 0.7$.

\section{Observations}
\J0006\ is an infrared-luminous quasar with faint ultraviolet (UV) emission.
It was first imaged by the Sloan Digital Sky Survey (SDSS; York et al. 2000) at the five optical wide-bands.
Meanwhile, the UKIRT Infrared Deep Sky Surveys (UKIDSS; Lawrence et al. 2007) and
the Wide-field Infrared Survey Explorer (WISE; Wright et al. 2010) provide
the NIR to middle-infrared (MIR) photometric data for \J0006.
The photometric data are summarized in Table \ref{tab1}.

The optical spectrum of \J0006\ was taken with the SDSS 2.5 m telescope on Dec. 16, 2011,
in the SDSS-III Baryon Oscillation Spectroscopic Survey (BOSS; Dawson et al. 2013)
and published in the SDSS Tenth Data Release (DR10; Ahn et al. 2013).
The BOSS spectrum has a wider wavelength range covering 361 nm - 1014 nm with a resolution of
1300 at the blue side and 2600 at the red side.
This instrument is fed by smaller optical fibers, each subtending 2'' on the sky (Smee et al. 2013).

The NIR spectrum of \J0006\ was performed with the TripleSpec spectrograph of the Hale 200-inch telescope (P200)
at Palomar Observatory on Dec. 4, 2015. Four exposures of 240 seconds each are taken in an A-B-B-A dithering model.
TripleSpec (Wilson et al. 2004) provides simultaneous wavelength coverage from 0.9 to 2.46 $\mu$m at a resolution of
1.4 - 2.9 \AA\ with two gaps at approximately 1.35 and 1.85 $\mu$m owing to the telluric absorption bands. 
Fortunately, the redshifted H$\beta$ and H$\alpha$ emission lines are detected
with the TripleSpec at $J-$ and $K_s-$band, respectively.
The raw data were processed using IDL-based Spextool software (Vacca et al. 2003; Cushing et al. 2004).

\section{Data Analysis and Results}

\subsection{Broadband Spectral Energy Distribution}

After correcting for the Galactic reddening of $E(B-V)=0.078$ mag (Schlegel et al. 1998),
we transformed the photometric data and the optical and NIR spectroscopic flux
into the rest-frame values with its emission redshift of $z=2.3174\pm0.0003$, 
which is carefully determined by fitting a Gaussian to the \lya\ narrow line.
The photometric and spectroscopic data are shown by red squares and black and blue curves in Figure 1.
Noting that, the absolute flux calibration is operated based on the SDSS and UKIDSS photometry for both optical and NIR spectra.
With the multi-wavelength spectroscopic and photometric data, we construct the broadband
SED of \J0006\ spanning from 1100 \AA\ to 6.65 $\mu$m in the rest frame. 
For comparison, the quasar composite combined from the SDSS DR7 quasar composite
($\lambda \le 3000$ \AA; Jiang et al. 2011) and the NIR template ($\lambda > 3000$ \AA; Glikman et al. 2006).
It is obvious that the observed SED of \J0006\ shows a very different shape from the quasar composite.
There is a striking ``V''-shape turning at $i$-band in the SED of \J0006,
which is likely to be caused by a type of excess broadband absorption or the spectral combination
of two different continuum slopes.

For the first case, the unusual shape is quite similar to the excess broadband absorption near 2250 \AA\ (EBBA)
in BAL quasars reported by Zhang et al. (2015a).
Therefore, the spectrum of \J0006\ is fit to search the quasar-associated 2175 \AA\ absorber
using the quasar composite reddened by a parameterized extinction curve (Fitzpatrick \& Massa 1990) in the quasar restframe,
where a Drude component represents the potential 2175 \AA\ bump (more details can be found in Section 2 of Zhang et al. 2015a).
For \J0006, the parameters of the Drude component are $c3=92.07\pm3.83$, $x_0=4.71\pm0.02$ $\mu$m$^{-1}$,
and $\gamma=5.32\pm0.26$ $\mu$m$^{-1}$. The bump strength is $A_{\rm bump}=\pi c3/(2\gamma)=27.16\pm1.76$ $\mu$m.
Apparently, the width ($\gamma$) and strength ($A_{\rm bump}$) of the bump deviate heavily from those of the bumps for
the Milky Way (Fitzpatrick \& Massa 1990), LMC (Gordon et al. 2003) curves,
and the quasar-associated 2175 \AA\ extinction curves (Jiang et al. 2011; Zhang et al. 2015a) (see Figure 2 of Zhang et al. 2015a).
Therefore, the broadband absorption  model is not preferred.

For the second case, the fibers feed the SDSS-III BOSS spectrograph subtend a diameter of 2'' on the sky,
and light from quasar pairs with small angular separations can be included through such a fixed aperture.
The physical size, to which the diameter corresponds, indicates that the spectrum of \J0006\ would also be polluted by its host galaxy. However,
the \lya\ and \ha\ emission lines simultaneously presented in the blue and red ends of the optical-NIR spectrum
suggest that both spectral components of \J0006\ could be quasar-type spectra.
Thus, we try the combination of a red quasar spectrum and a blue one,
which are the dominant components of the spectrum of \J0006\ with $\rm 3000~ \AA\ \lesssim \lambda \lesssim 7000$ \AA\ and $\lambda \lesssim 2000$ \AA, respectively.
Here, we use (1) a quasar composite scaled and multiplied by the SMC extinction law with a free parameter $E(B-V)$ to refer to the red component, (2) a quasar composite scaled by a power-law form ($\varpropto \lambda^{-\alpha_{\lambda}}$) to match the warped blueward.
In addition, a hot dust emission component is used to add the emission lack of the infrared composite ($\lambda \gtrsim 9000$ \AA).
Finally, we can decompose the broad-band SED in rest frame wavelength with the following model:
\begin{eqnarray}
F_{\lambda} = C_1 \lambda^{-\alpha_{\lambda}} F_{composite,\lambda} + C_2 A\left(E(B-V),\lambda\right) F_{composite,\lambda} +C_3 B_{\lambda}\left(T_{dust}\right),
\end{eqnarray}
where $C_1$, $C_2$ and $C_3$ are the factors for the respective components, $A\left(E(B-V),\lambda\right)$ is the dust extinction to the quasar emission, $F_{composite,\lambda}$ is the quasar composite, and $B_{\lambda}\left(T_{dust}\right)$ is the Planck function.
We perform least-squares minimization using the Interactive Data Language (IDL) procedure MPFIT developed
by C. Markwardt\footnote{The Markwardt IDL Library is available at http://cow.physics.wisc.edu/~craigm/idl/idl.html.}.
The values of $E(B-V)$ and $\alpha_{\lambda}$ are $0.7512\pm0.0049$ and $1.1388\pm0.0061$, respectively.
The typical temperature of hot dust is  $T\sim 1600$ K.
In Figure 1, the sum and three components of the best-fit model are shown by green and pink curves.

The fitting results suggest that the continuum of \J0006\ is heavily suppressed by dust reddening;
however, the warped blueward is even bluer than the average slope of quasars. 
We notice that the broadband SED of \J0006\ is similar to those of three high polarized quasars, i.e.,
PKS 2355-535 (Scarpa \& Falomo 1997), OI 287 (Li et al. 2015) and SDSS J091501.71+241812.1\footnote{
SDSS J0915+2418 was observed with the TripleSpec spectrograph of the Hale 200-inch telescope (P200)
at Palomar Observatory on 2014 January 18, and the SPOL CCD imaging/spectropolarimeter at the 6.5 m MMT on 2016 April 3.
The MMT/SPOL observation shows that the mean polarzed degree of the optical spectrum is $\sim15\%$ (C.-W. Yang, et al. 2016, in preparation).}.
This provides a guess that the warped blueward of \J0006\ is the scattered component arising from an extensive region.
In such a situation, the warped blueward would present high polarized fluxes;
unfortunately, the faint magnitudes of \J0006\ are unfit for the current spectropolarimetry observation.

\subsection{Emission-Line Spectrum}

The spectrum of \J0006\ shows strong \lya+\nv, \siiv+\oiv,\civ, \hb+\oiii\ and \ha\ emission lines;
however, ambiguous \ciii+\siii+\aliii, and \mgii\ emission lines are submerged in large noises.
 To further study the emission lines, we first subtract the underling pseudo continuum from the observed spectra.
In this work, linear functions ($f_{\lambda}=c_1*\lambda+c_2$) estimated from two continuum windows
are used as the local continua of the emission lines.
In \hb+\oiii\ and \ha\ regimes, we also adopt the I Zw 1 \ion{Fe}{2} template provided by V{\'e}ron-Cetty et al. (2004) and convolve it with a
Gaussian kernel in velocity space to match the width of \ion{Fe}{2} multiplets in the observed spectra.
The continuum windows are [1160, 1180] \AA\ and [1270, 1290] \AA\ for \lya+\nv,
[1320, 1350] \AA\ and [1430, 1450] \AA\ for \siiv+\oiv,
[1480, 1500] \AA\ and [1600, 1620] \AA\ for \civ, [4320, 4720] \AA\ and [5080, 5300] \AA\ for \hb+\oiii,
and [6150, 6250] \AA\ and [6850, 6950] \AA\ for \ha.
In the right panels of Figure 2, the local continua and \ion{Fe}{2} multiplets are marked by gray dashed and pink lines.
After subtracting the continuum model, we show the profiles of emission lines in their common velocity space in the left panels of Figure 2.

It is also clearly shown that the peaks of broad lines in \J0006\ tend to be blueshifted with respect to the \lya\ narrow line,
which indicates the profiles of these broad lines contains an emission-line outflow component. 
In particular, the line profiles of \civ\ and \lya+\nv\ of \J0006\ are almost the same as
those of IRAS 13224-3809 and 1H 0707-495 reported by Leighly et al. (2004a).
The profile of the \civ\ line is even entirely blueshifted (left-top panel);
 the unusual \civ/\ha\ width ratio, $\sim 0.71$, is larger than the value in the quasar composite spectrum (see Table 2).
Those suggest that the normal UV BELs of \J0006\ might also be obscured as the continuum.
If it is true, the \civ\ profile we observed is dominated by the emission-line outflows
and the gravitationally-bound component, i.e.,the normal BEL,  of \civ\ is possibly absence.
Here, we start the line profile fittings with the \civ\ line by one Gaussian.
The peak velocity of the \civ\ BBEL is blueshifted 2119 \kms\ and the full width at half maximum (FWHM) is 4821 \kms.
Later, we use the modelled profile of \civ\ line to recover the BBEL outflows of other emission lines.

 Firstly, the emission line of \siiv\ is depicted by the scaled \civ\ profile.
Actually, the strength of \siiv\ line is polluted by the emission of \oiv$\lambda\lambda\lambda\lambda\lambda$1397,1399,1401,1404,1407
(see Figure 7 of Vanden Berk et al. 2001).
A similar approach is taken in the fitting of the \lya\ and \nv\ lines,
where the normal BEL components of of which the normal BEL components are absent.
Then we use one narrow Gaussian component to model the \lya\ narrow line
and two broad Gaussian components to model the \lya\ and \nv\ BBEL components.
The shifts and widths of broad Gaussian components are tied to those of the \civ\ line.
In the right-top panel, the sum of the local continuum (gray dotted line) and the \lya\ narrow Gaussian (green line)
plus the  \lya\ and \nv\ BBEL outflows (blue lines) can well recover the observed spectrum.

Compared with UV emission lines, hydrogen Balmer lines in the optical waveband are still dominated by the gravitationally-bound BELRs,
and their profiles might only contain weak BBEL outflows.
To separate  the hydrogen Balmer emission-line outflows, we use three Gaussian profiles to model the entire profile of the \ha\ line.
One with the shift and width settings of the \civ\ line represents the BBEL outflows,
and the other two with the center wavelengths set to 6564.41 \AA\ represent the normal \ha\ BEL.
They are shown by blue and green lines in the right-middle panel, respectively.
In the \hb+\oiii\ regime, we use the unshifted two-Gaussian profile of the \ha\ line to model the normal \hb\ BEL (green line) and three Gaussians (blue lines) with the same shifts and widths as the \civ\ line to model the BBEL outflows of the \hb\ and \oiii-doublets in sequence.
The details of these components are shown in the right-bottom panel of Figure 2.

For all of the emission lines, we list the best-fit parameters in Table 2. Here, let us revert to the left panels of Figure 2.
We show the total fitted profile of each emission line by red lines and
the sum of the unblueshifted components of the corresponding line (and other lines only in the \lya+\nv\ or \hb+\oiii\ regimes) by pink lines.
The differences between the two above items are the BBEL outflows of each line (blue line).
The fit between the observed spectra with the continuum subtracted and the total fitted profiles
indicates the rationality of the existence of the BBEL outflows in all emission lines. 
Although the shifts of the peaks of the \ha\ line and the profile fitting of the \hb+\oiii\ regime strongly suggest the existence of hydrogen Balmer line outflows, it must be said that the BBEL strengths of hydrogen Balmer lines,
have large measurement uncertainty because of the multicomponent decomposition.

 The errors of line fluxes provided by MPFIT do not account for the uncertainty introduced by the pseudo continuum subtraction. To take this and other possible effects into account, we adopt a bootstrap approach to estimate the typical errors for our emission-line measurement. We generate 500 spectra by randomly combining the scaled model outflow emission lines (denoted as `A') and
the scaled model unblueshifted emission lines (denoted as `B'), to the scaled model pseudo continuum (denoted as `C').
Then, we fit the simulated spectra following the same procedure as described in above paragraphs.
For each parameter, we consider the error typical to be the standard deviation of the relative difference between the input ($x_i$) and the recovered ($x_o$), $\frac{x_o-x_i}{x_i}$. These relative differences turn out to be normally distributed for every parameter.
The thus estimated typical 1$-\sigma$ errors  are $\gtrsim  10\%$ for \civ, \lya\ and \nv, and  $\gtrsim  20\%$ for \ha, \hb\ and \oiii,
the derived flux errors are listed in Table 2.

Various interesting facts are summarized as follows:
(1) The \lya\ narrow line is the only narrow emission line in the spectrum of \J0006.
The $FWHM$ of the \lya\ narrow line is $429\pm30$ \kms, and the flux is $(20.81\pm0.89) \times10^{-17}$ erg s$^{-1}$.
(2) The extreme \civ\ line with $EW\rm_{C IV}=117.79\pm8.10~\AA$, far exceeds that of typical quasars ($EW\rm_{C IV}\approx20-50~\AA$).
\J0006\ and 12 other extremely red objects with $EW\rm_{C IV}\ge100~\AA$ are actually classified as Extreme EW objects, which
were considered to possibly be caused by suppressed continuum emission analogous to type II quasars in the unified model in Ross et al. (2015).
However, flux emitted from the abnormal BELR is the other important reason, and it is interpreted as the BBEL outflows in this work.
 (3) The \ha$/$\hb\ flux ratio of their unblueshifted components is $7.15\pm2.27$; however the intrinsic value of \ha$/$\hb\ of BELR is 3.06, with a standard deviation of 0.03 dex (Dong et al. 2008). The extremely high flux ratio suggests that the extinction of the BELR
is $E\rm(B-V) = 0.92^{+0.26}_{-0.34}$, which does not contradict the extinction we measured from the continuum of \J0006.
(4) However, the \ha$/$\hb\ flux  ratio of the outflowing components is $3.10\pm1.01$, which suggests that the outflow emission region is not obscured and reddened.

\section{Discussion}

\subsection{Central black hole and underrated luminosity }
From the emission line fittings, we can derive the central black hole mass using the commonly used virial mass estimators.
We use the broad \ha\ line-based mass formalism given by Greene \& Ho (2005):
\begin{eqnarray}
M_{\rm BH}=(2.0_{-0.3}^{+0.4}) \times 10^6 \left(\dfrac{L_{\rm H\alpha}}{10^{42}~{\rm erg~s^{-1}}}\right)^{0.55\pm0.02}
\left(\dfrac{FWHM_{\rm H\alpha}}{10^3~{\rm km~s^{-1}}}\right)^{2.06\pm0.06} M_{\sun},
\end{eqnarray}
where $L_{\rm H\alpha}$ and $FWHM_{\rm H\alpha}$ are the luminosity and width of \ha\ line.
Here, we only use the unblueshifted component emitted from the gravitationally-bound BELR.
Supposing that the broad emission lines also follow the continuum extinction,
the corrected luminosity of the normal \ha\ BEL is  \mbox{$L\rm_{H\alpha}=4.74\times10^{45} ~erg~s^{-1}$}.
 Together with \mbox{$FWHM\rm_{H\alpha}=6927~km~ s^{-1}$},
the central black hole mass is estimated to be $M\rm_{BH}=1.13\times10^{10}~\rm M_{\sun}$ with an uncertainty of a factor
of $\sim 2.6$ (from the intrinsic scatter $\sim 0.41$ dex for this single-epoch method compared with the
results of reverberation mapping, Ho \& Kim 2015).
The monochromatic continuum luminosity $L_{\rm 5100}=\lambda L_{\lambda}$ at 5100 \AA\
is directly calculated from the local continuum of the \hb\ regime,
and the extinction corrected luminosity $L_{\rm 5100}=6.46\times10^{46}$ \mbox{erg~ s$^{-1}$}.
The bolometric luminosity, $L_{\rm bol}=2.30\times10^{47}$ \mbox{erg s$^{-1}$},
is estimated from the monochromatic luminosity, $L_{\rm 5100}$, using the conversion given by Runnoe et al. (2012).
Our bootstrap approach showed the typical $1-\sigma$ error of the continuum under \hb+\oiii\ emission lines is $\sim 10\%$,
then the uncertainty of $L_{\rm bol}$ is also in the order of ten per cent.
The derived Eddington ratio is thus
$l_{\rm E}=L_{\rm bol}/L_{\rm Edd}=0.17$.
Based on the bolometric luminosity, the amount of mass being accreted is estimated as
$\dot{M}_{acc}=L\rm_{bol}/\eta c^2 =41~M_{\sun}~yr^{-1}$,
where we assumed an accretion efficiency $\eta$ of 0.1, and $c$ is the speed of light.

From the point of view of UV-band photometric and spectroscopic observations, \J0006\ is a fairly blue but faint quasar.
The relative color and  absolute magnitude at $i$-band are $\Delta(g-i)= -0.22$ and
$M_i = -24.17$, bluer and fainter than most quasars in the SDSS-III/BOSS survey.
If our inference regarding the unusual SED of \J0006\ is true, it is heavily obscured with $E(\rm B-V)=0.75$.
And the above-mentioned characters about color and luminosity are measured based on the scattering component.
The luminosity of \J0006\ is heavily underestimated;
when we compare the UV magnitudes and spectrum of \J0006\ with the quasar composite,
it is easy to see that the observed magnitudes of \J0006\ are fainter than the representation by approximately 4.5 mag.
Following the quasar luminosity functions measured in the SDSS-III/BOSS and SDSS-IV/eBOSS
(e.g, Ross et al. 2013; Palanque-Delabrouille et al. 2016),
we can derive the quasar number density with similar redshift and luminosity as follows:
$\varnothing (z\sim2.2-2.6, M_i\sim-28)\sim 10^{-8}\rm~Mpc^{-3}~mag^{-1}$.
The total spectroscopic footprint of the BOSS DR12 is pproximately 10,400 deg$^2$,
and the value without masked regions due to bright stars and data that do not meet the survey requirement is 9376 deg$^2$.
If there are quite a number ($\sim 100$) of quasars that have a similar SED and redshift  similar to those of \J0006,
that will redouble the bright-end of the quasar luminosity function
and lead to the statistical aberrations regarding the type-I and type-II ratio of the luminous AGNs at an intermediate redshift.
A statistical sample study in the future would be very useful to sort out this question.

\subsection{Physical properties of the outflows }

As described in \S 3.3, the decomposition of the emission line profiles indicates that
the presence of outflows in emission as revealed by the blueshifted components of
\civ, \lya, \nv, \oiii$\lambda\lambda$4959,5007, and hydrogen Balmer lines,
with a similar symmetric profile with the blueshifted velocity of 2119 \kms\ and the $FWHM$ of 4821 \kms.
In this subsection, we analyze and determine the physical properties of the emission-line outflows
using the photoionization synthesis code $Cloudy$ (the latest version, last described by Ferland et al. 1998).

To investigate the physical properties for the emission-line outflows, we use the $Cloudy$ simulations
and confront these models with the measured line ratios to determine the density ($n_H$),
the column density ($N_H$) and the ionization parameter ($U$).
 We know that the BBEL manifests itself well in \civ, \siiv+\oiv, \lya, and \nv\ with the entire emission-line profile being blueshifted since the normal BELs are heavily suppressed, and the BBEL also dominates the emission of \oiii.
However, \siiv+\oiv\ BBEL is weak with large uncertainty.
Moreover, the resonant scattering can contribute a significant part of the \nv\ line and it may give rise to anomalous strong \nv\ line
(Hamann, Korista \& Morris 1993; Wang et al. 2007).
Even in some objects, e.g., 0105-265, almost half of the \lya\ fluxes are scattered (Ogle et al. 1999).
Thus, \lya, \civ\ and \oiii\ are chosen to confront the models,
and \siiv+\oiv\ and the absent \ciii\  are just used to examine the reasonability of the constrained parameter space.
We consider that the BBEL emission region is not obscured, and adjacent lines are not urged.
It is better for us to use the line ratios with large differences in ionization potential,
i.e., \oiii$\lambda$5007/\lya\ and \civ/\lya, in order to probe different zone in a gas cloud.
Furthermore, the critical density ($n_{\rm crit}$) of \oiii$\lambda\lambda$4959,5007 is $7.0\times10^5\rm~cm^{-3}$,
and we, thus, infer that the BBEL outflows should originate from the less dense part of the outflows, with a density lower than $n_{\rm crit}$.

We consider a gas slab illuminated by a continuum source in the extensive parameter space.
The outflowing gases are assumed to have uniform density and a homogeneous chemical composition of solar values and be free of dust.
The incident SED applied is a typical AGN multi-component continuum described as a combination of a blackbody ``Big Bump''
and power laws\footnote{see details in Hazy, a brief introduction to $Cloudy$; http://www.nublado.org}.
The ``Big Bump" component peaks at $\approx$ 1 Ryd and is parameterized by T=$1.5\times10^5$ K.
The slope of the X-ray component, the X-ray to UV ratio, and the low-energy slope are
set to be $\alpha_{\rm x}=-1$,  $\alpha_{\rm ox}=-1.4$, and $\alpha_{\rm UV}=-0.5$, respectively.
 This UV-soft SED is regarded more realistic for radio-quiet quasars than the other available
SEDs provided by $Cloudy$ (see the detailed discussion in \S4.2 of Dunn et al. (2010)).
We calculated a series of photoionization models with different ionization parameters, electron densities and
hydrogen column densities. The ranges of parameters are $-3\leqslant{\rm log_{10}}~U\leqslant1$,
$1\leqslant{\rm log_{10}}~n_{\rm H}~({\rm cm^{-3}})\leqslant6$ and
$18\leqslant{\rm log_{10}}~N_{\rm H}~({\rm cm^{-2}})\leqslant24$ with a step of 0.2 dex.

 We extract the simulated \oiii$\lambda$5007, \civ\, and \lya\ fluxes from the Cloudy simulations
and show the line ratios  \oiii$\lambda$5007/\lya\ and \civ/\lya\ by the red and green lines in Figure 3.
The red and green areas show the observed $1-\sigma$ uncertainty ranges of \oiii$\lambda$5007/\lya\ and \civ/\lya,
so the overlapping region is the possible parameter space for the BBEL outflows of this object.
Model calculations suggest that the gas clouds in two irregular regions with ${\rm log_{10}}~N_{\rm H}~({\rm cm^{-2}})\sim 21.0$ and
${\rm log_{10}}~N_{\rm H}~({\rm cm^{-2}})\sim 22.0$ can generate the observed line ratios of \J0006 (Figure 3).
Fortunately, the absence of \ciii\ gives the upper limit of $9.71\times10^{-17}\rm erg~s^{-1}~cm^{-2}$, the flux ratio of \ciii/\lya\ (cyan area in Figure 4) just cover the high column density region as the reliable parameter space.
It is regrettable that the density cannot be more firmly confined.
The $Cloudy$ simulation gives a density range of ${\rm log_{10}}~n_{\rm H}~({\rm cm^{-3}})\leqslant5.8$,
which is an upper limit, as \oiii$\lambda\lambda$4959,5007 also suggested.
In Figure 5, the modeled and observed line ratios of \siiv+\oiv\ and \lya\ are shown in blue.
It is clear that observed \siiv+\oiv/\lya\ is overlapped with the \oiii$\lambda$5007/\lya\ and \civ/\lya\ areas,
but \siiv+\oiv/\lya\ makes the possible parameter space of the outflowing gases narrower.
In summary, the outflowing gases can survive in the parameter space with ${\rm log_{10}}~n_{\rm H}~({\rm cm^{-3}})\leqslant5.8$, ${\rm log_{10}}~U \sim -0.5$, and ${\rm log_{10}}~N_{\rm H}~({\rm cm^{-2}})\sim$ 22.0. Actually, the reliable parameter-space region is essentially dynamic with the gas density.

\subsection{Kinetic luminosity and mass flux of the outflows}
Based on the physical properties determined by $Cloudy$, we estimate the distance $R$ of the outflows away from the central source.
The ionization parameter $U$ depends on $R$ and the rate of hydrogen-ionizing photons emitted by the central source $Q$
and is given by
\begin{eqnarray}
U = \int^{inf}_{\nu_0} \dfrac{L_{\nu}}{4\pi R^2 h\nu n_{\rm H}c}~d\nu=\dfrac{Q}{4\pi R^2 n_{\rm H}c},
\end{eqnarray}
in which,  $\nu_0$ is the frequency corresponding the hydrogen edge, $n_{\rm H}$ is is the density of the outflows and $c$ is the speed of light.
To determine the hydrogen-ionizing rate $Q$, we scale the UV-soft SED to the extinction corrected flux of \J0006\ at 5100\AA\ (rest-frame)
and then integrate over the energy range $h\mu\ge 13.6$ eV of this scaled SED. This yields $Q = 1.17 \times 10^{57}~\rm photons~s^{-1}$.
$n_{\rm H}$ has been estimated as $n_{\rm H}\leqslant10^{5.8}~\rm cm^{-3}$.
Using this $Q$ value together with the derived $n_{\rm H}$ and $U$, the lower limit of $R$ can be derived to as 41 pc.
 The radius of broad emission-line regions, $R_{\rm BLR}$,
can be estimated using the formula based on the luminosity at 5100\AA,
\begin{eqnarray}
R_{\rm BELR} = \alpha \left(\dfrac{L_{5100}}{10^{44}~\rm erg~s^{-1}}\right)^{\beta}~{\rm lt-days,}
\label{rblr}
\end{eqnarray}
where the parameters, $\alpha$ and $\beta$ are $30.2\pm1.4$ and $0.64\pm0.02$ given in Greene \& Ho (2005).
Thus, the luminosity yields $R_{\rm BLR}\approx1.3$ pc.
Meanwhile, the radius of the inner side of the dust torus, $R_{\rm Torus}$, can also be estimated based on the thermal equilibrium of the inner side of the torus as
\begin{eqnarray}
R_{\rm Torus}=\sqrt{\dfrac{L_{\rm bol}}{4\pi\sigma T^4}},
\end{eqnarray}
where $\sigma$ is the Stefan-Boltzmann constant, $T (\sim 1500~\rm K)$ is the temperature of inner side of the tours.
Then, we get $R_{\rm Torus}\approx 16$ pc.
Formulas in Burtscher et al. (2013) and Kishimoto et al. (2011) gave the dust sublimation radii as 5.4 pc and 2.0 pc respectively, smaller than the above estimation. In any case, the outflows are located externally of the torus.
This result is consistent with the supposition of our spectral qualitative analysis, i.e., the dusty torus only obscures the continuum and the normal BELR rather than the BBEL outflows.

Assuming that the outflow materials can be described as a thin partially filled shell,
the average mass-flow rate ($\dot{M}$) and kinetic luminosity ($\dot{E}_k$) can be derived as follows (Borguet et al. (2012)),
\begin{eqnarray}
\dot{M}=4\pi R\Omega\mu m_p N_H v,
\end{eqnarray}
\begin{eqnarray}
\dot{E}_k=2\pi R\Omega\mu m_p N_H v^3,
\end{eqnarray}
where $R$ is the distance of the outflows from the central source, $\Omega$ is the global covering fraction of the outflows,
$\mu = 1.4$ is the mean atomic mass per proton, $m_p$ is the mass of proton,
$N_H$ is the total hydrogen column density directly derived from the photoionization modeling of the outflow gases,
and $v$ is the flux weight-averaged velocity of the outflow gases. Here, $v$ is replaced by the blueshifted velocity of the BBEL peaks.
We note that the BBELs originate from the gases outflowing along different directions with respect to the observer,
thus the observed outflow velocity of the BBEL is a sum of the projected velocities of the outflowing gases along different directions,
and should be a lower limit of the velocity of the outflow materials.
Furthermore, we can estimate the global covering fraction for \J0006\ by comparing the measured $L_{\rm \oiii\lambda 5007}$
with the predicted one by the $Cloudy$ model.
In $Cloudy$ modeling, the emergent values of  $L_{\rm \oiii\lambda 5007}$ are output with the covering fraction being assumed to be 0.05 to 1
and are shown  as a function of $n_H$ ($U$) and $\Omega$ in Figure 6. Meanwhile, we show the measured $L_{\rm \oiii\lambda 5007}$ with errors in gray.
The comparation gives the global covering fraction for \J0006\ to be $\sim 6\%-15\%$.
In the studies of BAL quasars, the  fraction of BAL quasars is generally used to be as the global covering fraction of BAL outflow gases.
The fraction is generally $10\%-20\%$ in optical-selected quasars (e.g., Trump et al. 2006; Gibson et al. 2009; Zhang et al. 2014),
$\sim 30\%$ (e.g., Maddox \& Hewett 2008) or even 44\% (Dai et al. 2008)
in NIR-selected quasars. We consider that BAL and BEL outflows may be different appearances viewed from different inclination angles
but represent the same physical component which is suggested by many works (e.g., Richards et al. 2011; Wang et al. 2011; Liu et al. 2016),
the global covering fraction of BEL outflows will follow these values.
The value we obtained in \J0006\ is slightly smaller than those estimated from optical-selected BAL quasars.
Thus, for the upper limit of the density $n_H=10^{5.8}~\rm (cm^{-3})$ and the  global covering fraction  $\sim 10\%$,
the kinetic luminosity and mass loss rate are calculated as
$\dot{E_k}=1.74\times10^{43}~\rm erg~s^{-1}$,
$\dot{M}=12.23~\rm M_{\sun}~yr^{-1}$ for $U=10^{-0.5}~{,}~N_H=10^{22.0}~\rm cm^{-3}$

In general, the high-velocity outflows require kinetic luminosities to be of the order of
a few percent of the Eddington luminosity (e.g., Scannapieco \& Oh 2004; Di Matteo et al. 2005; Hopkins \& Elvis 2010);
however, in the case of \J0006, the reported lower limit on the kinetic luminosity ($\dot{E_k}\sim10^{-5}L_{Edd}$)
is not of a sufficient value to efficiently drive AGN feedback.
Note that this comparison is probably only a lower limit for the following reasons:
(1) The weight-averaged velocity $v$ we use in Formula 5 is a sum of the projected velocities of the outflowing gases along different directions,
the value of which is underestimated and should be a lower limit of the outflow velocity.
(2) The distance of the outflows, $R$, is also a lower limit, and the outflow gases perhaps survive at greater distances from the central source.

\subsection{Origin of the \lya\ narrow line }
As the only narrow emission line in \J0006, the \lya\ narrow line presents strong emission.
In theory \lya\ photons scatter in neutral hydrogen until they either escape or are absorbed by dust grains.
For \J0006, possible explanations are the NELR emission, the star formation in host galaxy,
or the scattering of the obscured \lya\ emission by the outflows. Before proceeding further,
we first discuss whether the NELR and star-forming contribution are preferred in \J0006\ through other relevant narrow lines.
If \lya\ of \J0006\ is the narrow emission line from the NELR, this situation can be confirmed through the presence of high-ionization lines,
e.g., \civ\ in the rest-frame UV, or other lines such as \ion{He}{2} which may also indicate AGN activity.
The typical relative intensities of \lya\ and \civ\ narrow emission lines in Seyfert 2 spectra are 55 and 12,
respectively (from Ferland \& Osterbrock 1986).
Under this condition, the derived flux of the \civ\ narrow line is $\rm \sim 4.54\times10^{-17} ~erg~s^{-1}cm^{-2}$.
Like the \lya\ narrow line, we use one Gaussian with a width of 429 \kms\ to model the theoretical \civ\ narrow line,
which is shown via the green line in the top panel of Figure 7.
Apparently, the theoretical the \civ\ narrow line would not appear in the spectrum of \J0006; otherwise,
it should be detected at $\gtrsim 7\sigma$.

With the assumption that the \lya\ emission is produced by photoionization by stars and the absence of extinction,
one can estimate the \ha\ flux while assuming a case B recombination theory (Osterbrock 1989),
$f_{esc}(Ly\alpha)=f(Ly\alpha)/(8.7\times f(H\alpha))$, where $f_{esc}(Ly\alpha)$ is the \lya\ escape fraction,
$f(Ly\alpha)$ is the observed \lya\ flux and $f(H\alpha)$ is the \ha\ flux. When \lya\ photons are 100\% escaped,
we obtain the  flux and luminosity of \ha\ for the \ion{H}{2} region, $f(H\alpha)=(2.39\pm0.10)\times10^{-17}$ \mbox{erg s$^{-1}$ cm$^{-2}$}
and $L\rm_{H\alpha}=(3.20\pm0.13)\times10^{42}~erg~s^{-1}$.
Following the relation star formation rate $SFR(M_{\sun}~yr^{-1})= 7.9\times10^{-42}~L_{H\alpha}~\rm (erg~ s^{-1})$ (Kennicutt 1998),
$SFR$ is estimated to be $SFR\rm = 25.28\pm 1.03 ~M_{\sun}~yr^{-1}$.
When we use the profile of the \lya\ narrow line to model the theoretical \ha\ (pink line in the bottom panel of Figure 7),
the height of this Gaussian profile is 0.26 \mbox{erg s$^{-1}$ cm$^{-2}$ \AA$^{-1}$},
which is commensurate with the  fluctuations of the NIR spectrum.
Previous studies find that $f_{esc}(Ly\alpha)$ is not constant and spans a wide range of values.
At a similar redshift to \J0006, the average escape fraction from various methods are
$5.3\pm3.8$ per cent by performing a blind narrowband survey in \lya\ and \ha\ (Hayes et al. 2010),
29 per cent at $1.9\le z \le 3.5$ with the comparison between \lya\ flux and dust-corrected UV continuum (Blanc et al. 2011),
$> 7$ per cent at $z=2.1$ and 3.1 with the comparison between \lya\ flux and X-ray flux (Zheng et al. 2012),
and $\sim 12-30$ per cent at $z=2.2$  with the comparison between \lya\ and \ha\ luminosity (Nakajima et al. 2012).
In the bottom panel of Figure 7, we also show the theoretical \ha\ profiles with $f_{esc}(Ly\alpha)=90\%$, 80\%, 70\%, 50\%, 30\%, and 10\% in sequence.
It becomes apparent that there is a threshold almost no possibility of the existence of the \ha\ narrow line superposed on the \ha\ narrow line,
however, it also cannot be ruled out by the low spectrum quality of the \ha\ regime of \J0006.
 Thus, the scattering of the obscured \lya\ emission by the outflows is the most promising option.

Alternately, the NELR in AGNs is the largest spatial scale for the ionizing radiation from the central source. 
In the case of the NELR, the \lya\ emission comes from a spatially extended region, so at least some extent
physical and kinematic distributions can be mapped out directly.
In comparison to the NELR, the star-forming region in galaxies is generally of compact morphology
and typically characterized by a small but resolved ``core'' with kpc scales rather the axisymmetric bi-conic ionized region.
The high-resolution \lya\ imaging will also clearly demonstrate the presence or location of the extent structures.
For the possible participation of scattering, spectropolarimetric observations of the \lya\ line will ask for final clarification.

\section{Conclusion }
We performed a multiwavelength study of the continuum and emission lines of \J0006, a high redshift quasar with striking ``V''-shape SED turning at $i$-band and signatures of outflows in blueshifted emission lines.
The multiwavelength SED analysis indicates that the continuum of \J0006\ is heavily suppressed by dust reddening with $E(B-V)=0.7512\pm0.0049$,
reducing the high luminosity of $L_{\rm bol}=2.63\times10^{47}~\rm erg~s^{-1}$ to $M_i=24.17$,
and the detected UV radiation comes from scattering. 
In the optical and NIR spectra, the normal BELR-originated emission lines, and the continuum, are obscured,
and, thus, the emission-line outflows are significantly presented, even the emission profiles of \civ, \lya\ and \nv\ are totally dominated
by the blueshifted components, suggestive of the AGN outflows. 
These lines can be used to constrain the physical properties of the outflowing gases by comparing the observed
results with the model results of the photoionization synthesis code $Cloudy$.
The physical parameters we determined for emission-line outflows are
$U\sim 10^{  -0.5}$ and $N_H\sim 10^{ 22.0}\rm ~cm^{-2}$,
and the density is regrettable given an upper limit of $n_H\sim10^{5.8}\rm~cm^{-3}$.
Using the luminosity of \oiii$\lambda$5007 to obtain the global covering factor and the upper limit of the density,
we estimate the lower limits on the distance of the outflow materials to the central source, 
the kinetic luminosity and mass loss rate of the outflows, which are $R\sim 41$ pc,
$\dot{E_k}=1.74\times10^{43}~\rm erg~s^{-1}$, and $\dot{M}=12.23~\rm M_{\sun}~yr^{-1}$, respectively.
Under conditions of the NELR emission and star formation in the host galaxy, the strengths of the \civ\ and \ha\ narrow lines are investigated based on the relations with the \lya\ narrow line.
The non-detection of the \civ\ and \ha\ narrow lines suggests the \lya\ narrow line of \J0006\ is not likely
to be originated in the NELR or the starforming region in the host galaxy.
Perhaps the outflow gases, which have expanded to the inner region of the NELR,
finally diffuse in the larger region and ace as the scatterers of \lya\ photons.

\acknowledgments{This work is supported by National Natural Science Foundation of China (NSFC-11573024, 11473025, 11421303) and  National Basic Research Program of China (the 973 Program 2013CB834905). T. Ji is supported by National Natural Science Foundation of China  (NSFC-11503022) and Natural Science Foundation of Shanghai (NO. 15ZR1444200). P. Jiang is supported by  National Natural Science Foundation of China  (NSFC-11233002).
We acknowledge the use of the Hale 200-inch Telescope at Palomar Observatory through the Telescope Access Program (TAP), as well as the archive data from the SDSS, UKIDSS and WISE Surveys. TAP is funded by the Strategic Priority Research Program. The Emergence of Cosmological Structures (XDB09000000), National Asto nomical Observatories, Chinese Academy of Sciences, and the Special Fund for Astronomy from the Ministry of Finance. Observations obtained with the Hale Telescope at Palomar Observatory were obtained as part of an agreement between the National Astronomical Observatories, Chinese Academy of Sciences, and the California Institute of Technology. Funding for SDSS-III has been provided by the Alfred P. Sloan Foundation, the Participating Institutions, the National Science Foundation, and the U.S. Department of Energy Office of Science. The SDSS-III Web site is http:// www.sdss3.org/.}

\clearpage

\begin{deluxetable}{ccccc}
\tabletypesize{\scriptsize}
\tablewidth{0pt}
\tablenum{1}
\tablecaption{Photometric Observations of SDSS J0006+1215
\label{tab1} }
\tablehead{
\colhead{Band}  & \colhead{Magnitude} & \colhead{Date of Observation}& \colhead{Facility} & \colhead{Reference}}
\startdata
$\textit{u}$          &$ 24.53\pm 1.13 $ &2008 Nov. 02&SDSS&1\\
$\textit{g}$          &$ 22.13\pm 0.09 $ &2008 Nov. 02&SDSS&1\\
$\textit{r}$          &$ 22.06\pm 0.13 $ &2008 Nov. 02&SDSS&1\\
$\textit{i}$          &$ 22.11\pm 0.18 $ &2008 Nov. 02&SDSS&1\\
$\textit{z}$          &$ 20.70\pm 0.23 $ &2008 Nov. 02&SDSS&1\\
$  Y  $               &$ 20.33\pm 0.28 $ &2010 July 26&UKIDSS&2\\
$  J  $               &$ 18.97\pm 0.11 $ &2010 July 26&UKIDSS&2\\
$  H  $               &$ 17.37\pm 0.05 $ &2010 July 26&UKIDSS&2\\
$  K  $               &$ 15.81\pm 0.03 $ &2010 July 26&UKIDSS&2\\
$  W1 $               &$ 14.32\pm 0.03 $ &2010 June 28&WISE&3\\
$  W2 $               &$ 12.74\pm 0.03 $ &2010 June 28&WISE&3\\
$  W3 $               &$  8.85\pm 0.03 $ &2010 June 28&WISE&3\\
$  W4 $               &$  6.55\pm 0.07 $ &2010 June 28&WISE&3
\enddata
\tablenotetext{References:}{(1) Ahn et al. (2013); (2) Lawrence et al. (2007); (3) Wright et al. (2010)}
\end{deluxetable}

\begin{deluxetable}{cccc  cccc}
\tabletypesize{\scriptsize}
\tablewidth{0pt}
\tablenum{2}
\tablecaption{Emission-Line Parameters of SDSS J0006+1215 \label{tab2} }
\tablehead{
\colhead{}  & \colhead{} & \multicolumn{4}{c}{SDSS J0006+1215} & \colhead{} &\colhead{Vanden Berk et al. composite}\\
\cline{3-6}
\colhead{Ions}  & \colhead{$\lambda_{\rm lab}$} & \colhead{Velocity}& \colhead{FWHM}  & \colhead{Flux}   & \colhead{Note}& \colhead{}&\colhead{Rel. Flux}  \\
\colhead{}  & \colhead{\AA}    & \colhead{(km s$^{-1}$)}& \colhead{(km s$^{-1}$)} & \colhead{($10^{-17}$ erg s$^{-1}$ cm$^{-2}$)} & \colhead{}&\colhead{}&\colhead{(100$\times F/F_{Ly\alpha}$)}
}
\startdata
Ly$\alpha$	&1215.67&0 	   &429$\pm$30   &20.81$\pm$1.91   & narrow & & \\
Ly$\alpha$	&1215.67&-2119 &4821 		 &105.55$\pm$12.20 & outflow& & 100.00 \\
N V 		&1240.14&-2119 &4821 		 &96.94$\pm$11.40  & outflow& &2.46 \\
Si IV+O IV] &1396.76&-2119 &4821         &13.00$\pm$4.32   & outflow& &8.13 \\
C IV 		&1549.06&-2119$\pm$179 &4821$\pm$486 &54.07$\pm$6.31  & outflow&  &25.29 \\
C III]      &1908.73&-2119 &4821         &9.71             & outflow, upper limit& &21.19 \\
H$\beta$ 	&4862.68&      &             &89.50            & total&  &8.65 \\
-	        &&0     &6927         &69.07$\pm$15.04  & broad&  &  \\
-        	&&-2119 &4821         &20.43$\pm$4.15   & outflow&  &  \\
$[$O III$]$ &4960.30&-2119 &4821     &15.01  		   & outflow&  &0.69 \\
$[$O III$]$ &5008.24&-2119 &4821     &45.04$\pm$9.03   & outflow&  &2.49 \\
H$\alpha$   &6564.61&      &             &557.45           & total&  &30.83 \\
-           &&0     &6927$\pm$231 &494.06$\pm$114.76& broad&  &  \\
-           &&-2119 &4821		 &63.39$\pm$16.34  & outflow&  &
\enddata
\end{deluxetable}

\figurenum{1}
\begin{figure*}[tbp]
\epsscale{1.0} \plotone{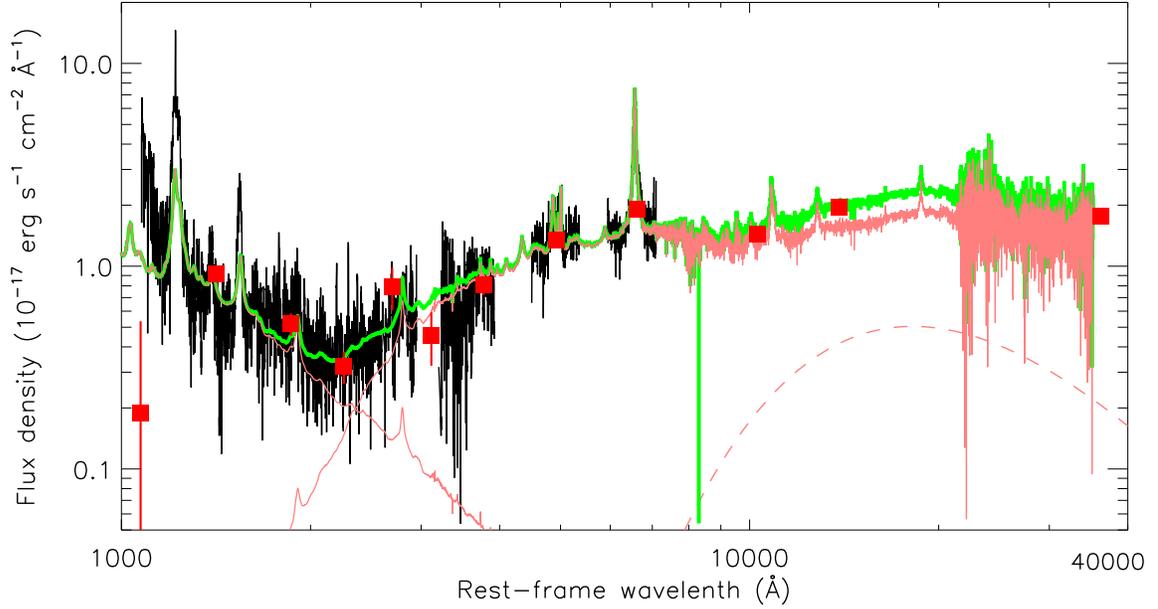}
\caption{Broadband SED of SDSS~J0006+1215 from UV to MIR by red squares, the spectra of SDSS and TripleSpec by black curves.
The scattering light, the scaled and reddened quasar composite, the hot dust emission and their sum are shown by pink and green curves. }\label{f1}
\end{figure*}

\figurenum{2}
\begin{figure*}[tbp]
\epsscale{0.9} \plotone{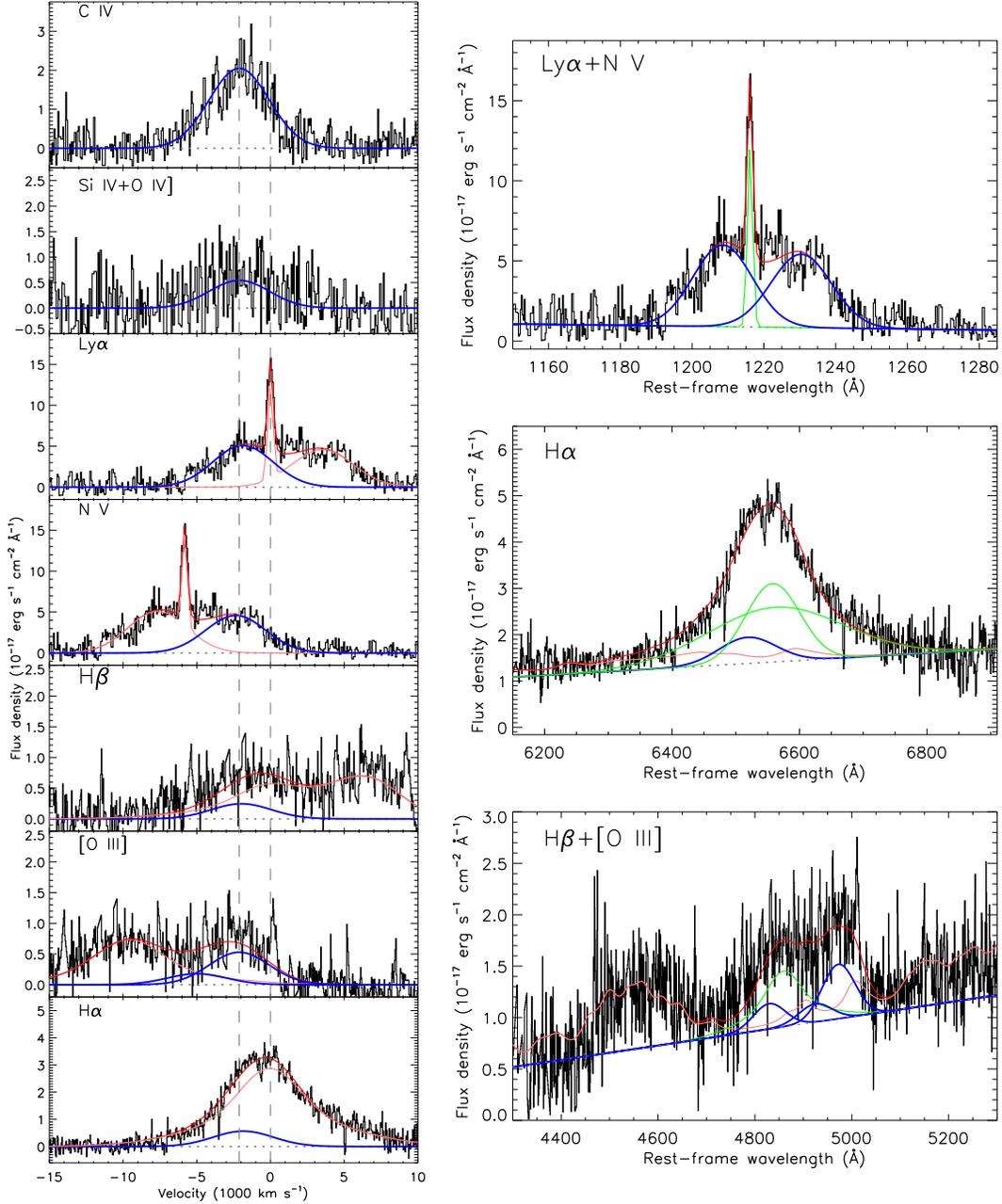}
\caption{Left panels: Demonstration of the profile of the various emission lines, particularly the blueshifted components caused by outflows.
Plotted are the observed spectra with continuum subtracted (black lines), the total fitted profile of each emission line (red lines),
the sum of the unblueshifted components of corresponding line and other lines (only in the \lya+\nv\ or \hb+\oiii\ regimes) (pink lines),
and broad blueshifted component (blue lines). The gray dashed vertical lines indicate $-2119$ \kms\  and 0 \kms.
Right panels: Best-fit models for the \lya+\nv, \ha, and \hb+\oiii\ regions. The red lines represent the total fit profile for each region.
The blue lines represent the blueshifted components, the green lines show the decomposed Gaussian profiles of the broad and narrow emission lines respectively, \textbf{the broadened optical \ion{Fe}{2} template in pink,}
and the gray dashed lines represent the local continua. 
}\label{f2}
\end{figure*}

\figurenum{3}
\begin{figure*} [htbp]
\rotatebox{270}{\includegraphics[width=0.70\textwidth]{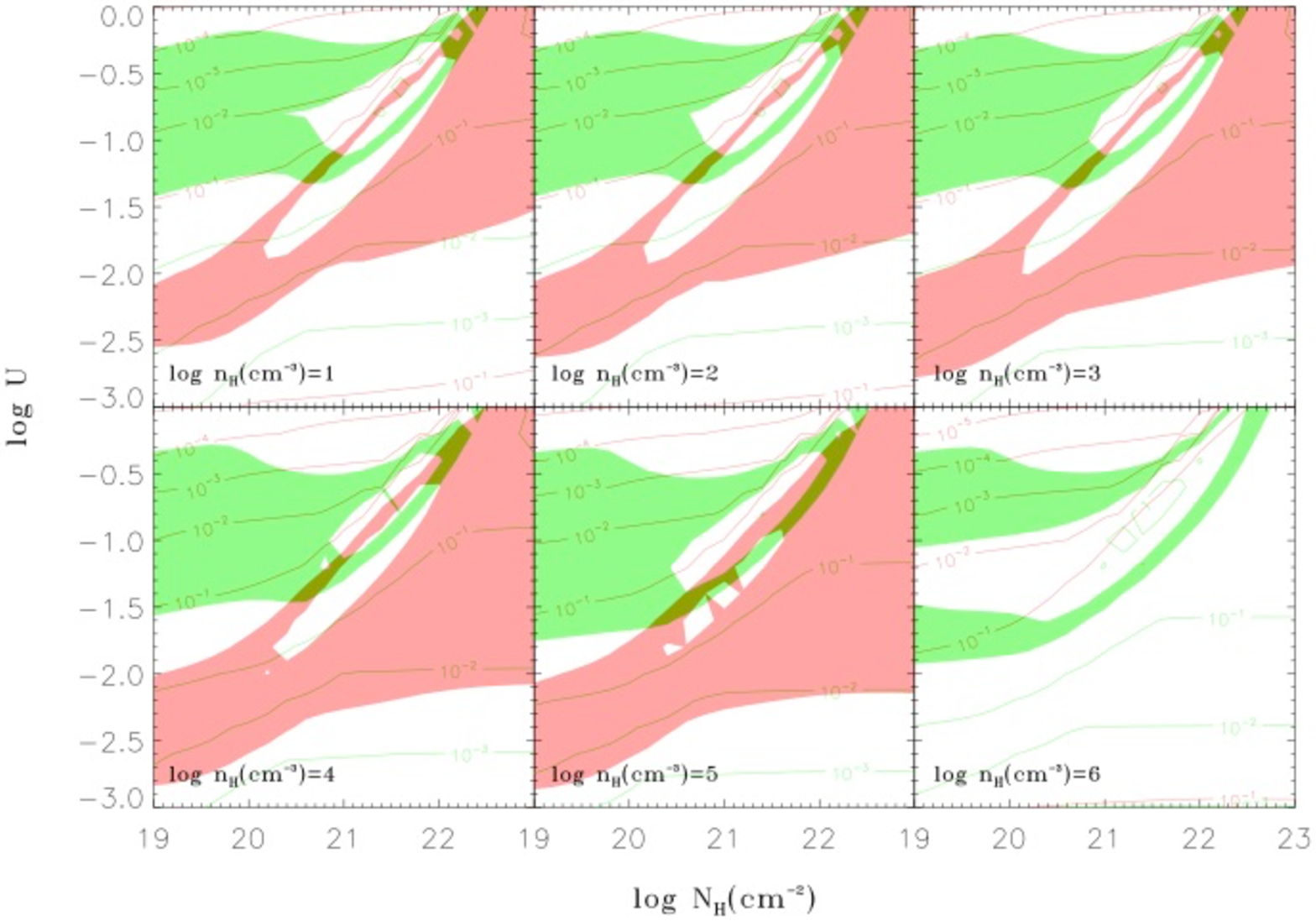}}
\caption{Photoionization models of the emission-line outflows in \J0006\ assuming solar abundances.
Red and green dashed lines represent the contours of flux ratios of \oiii$\lambda$5007/\lya\ and \civ/\lya, respectively.
The red and green areas show the observed $1-\sigma$ uncertainty ranges of \oiii$\lambda$5007/\lya\ and \civ/\lya,
the overlapping region is the possible parameter space for the BBEL outflows of this object.
}\label{f3}
\end{figure*}

\figurenum{4}
\begin{figure*} [htbp]
\rotatebox{270}{\includegraphics[width=0.70\textwidth]{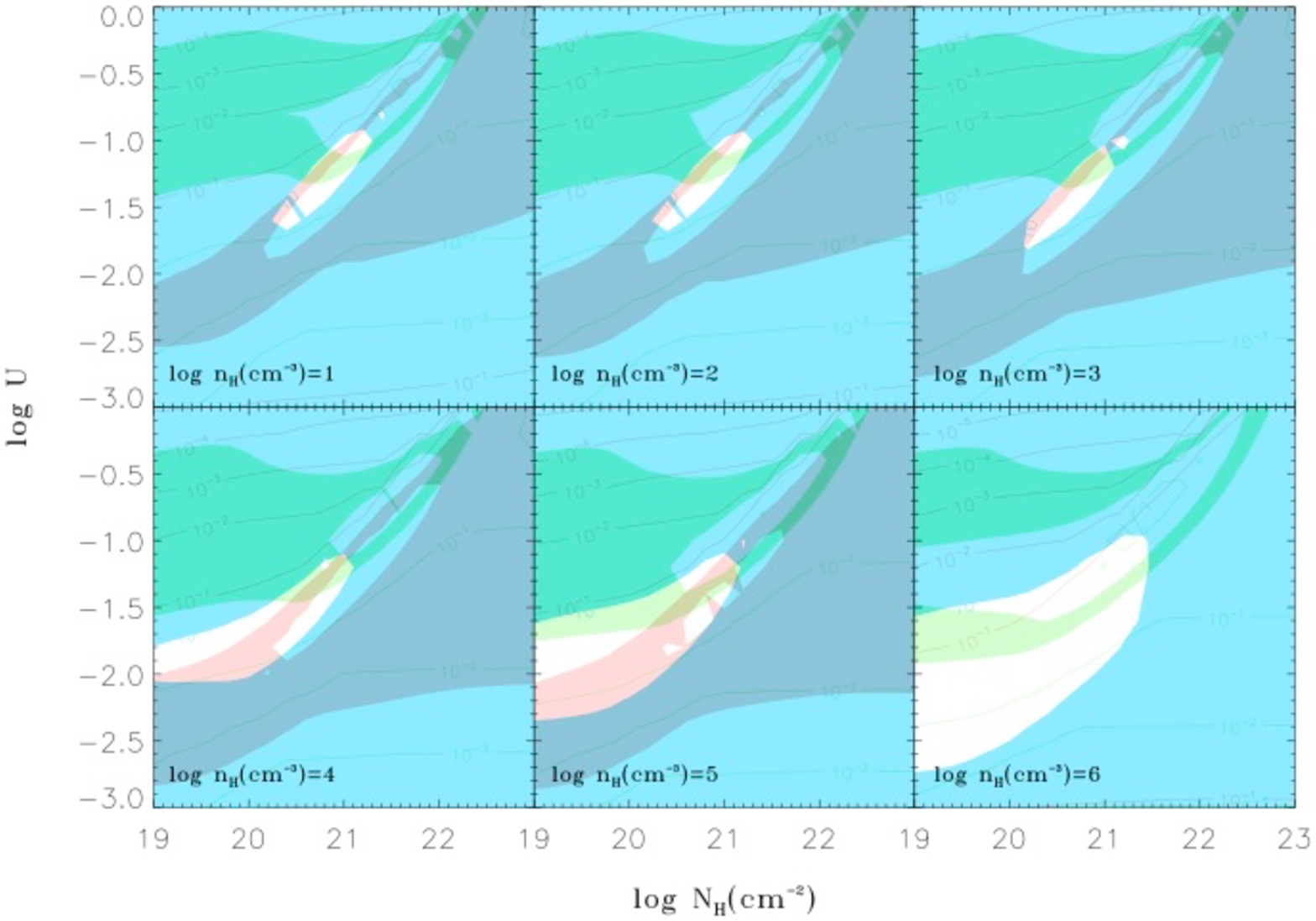}}
\caption{The upper limit of \ciii/lya\ overplots the results of \oiii$\lambda$5007/\lya\ and \civ/\lya\ in cyan.
}\label{f4}
\end{figure*}

\figurenum{5}
\begin{figure*}  [htbp]
\rotatebox{270}{\includegraphics[width=0.70\textwidth]{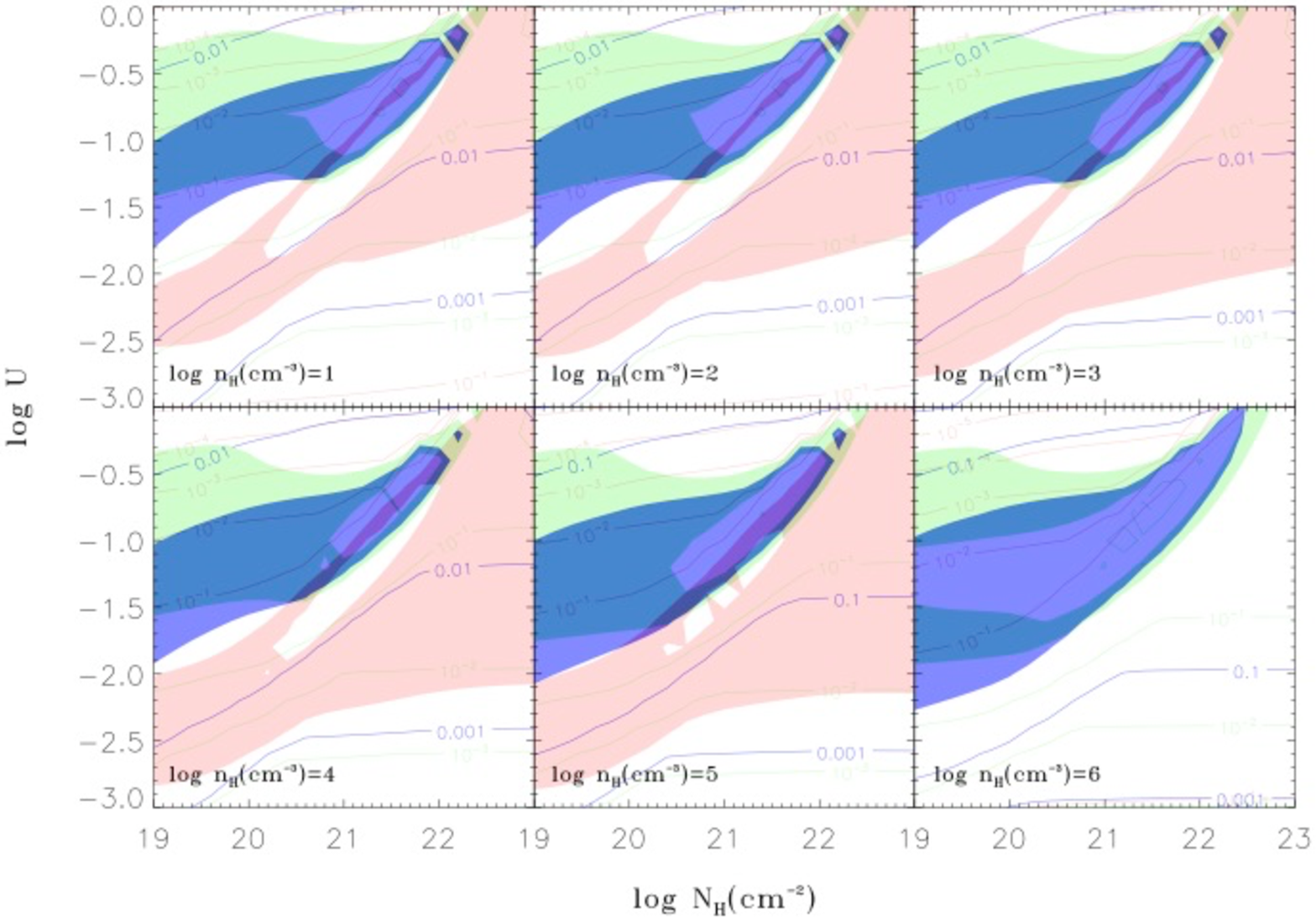}}
\caption{The flux ratios of \siiv+\oiv/\lya\ overplots the results of \oiii$\lambda$5007/\lya\ and \civ/\lya\ in blue.
}\label{f5}
\end{figure*}

\figurenum{6}
\begin{figure*}[tbp]
\epsscale{0.8} \plotone{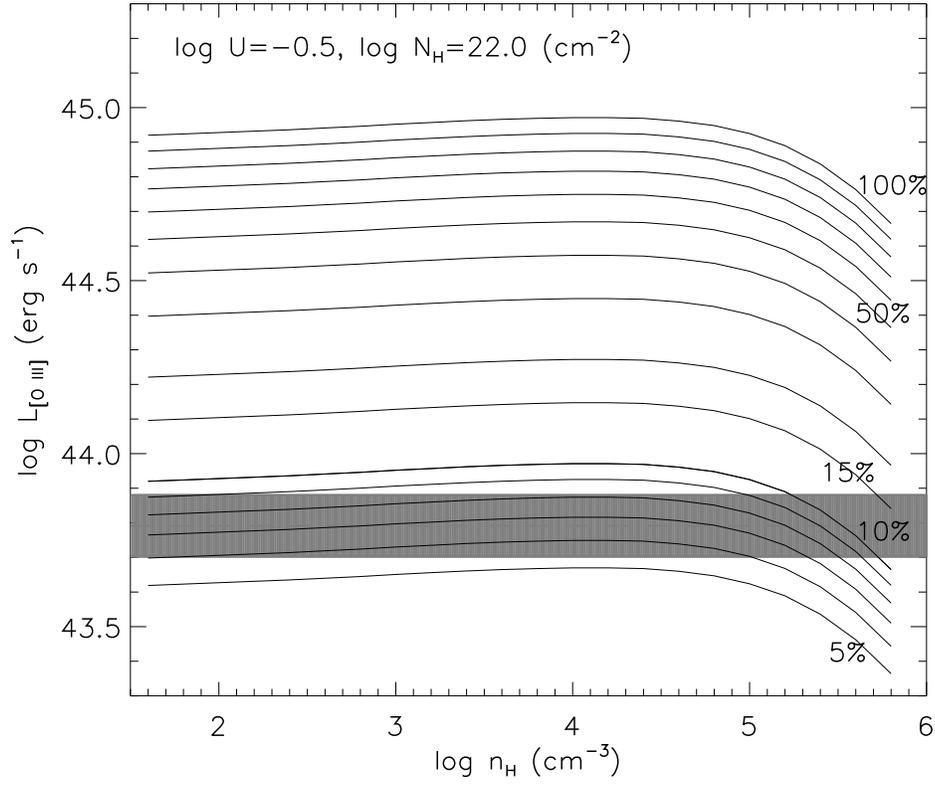}
\caption{Luminosity of \oiii$\lambda$5007 as a function of gas density  and global covering fraction,
the measured $L_{\rm \oiii\lambda5007}$ with $1-\sigma$ uncertainty is shown in gray.
}\label{f6}
\end{figure*}

\figurenum{7}
\begin{figure*}[tbp]
\epsscale{0.6} \plotone{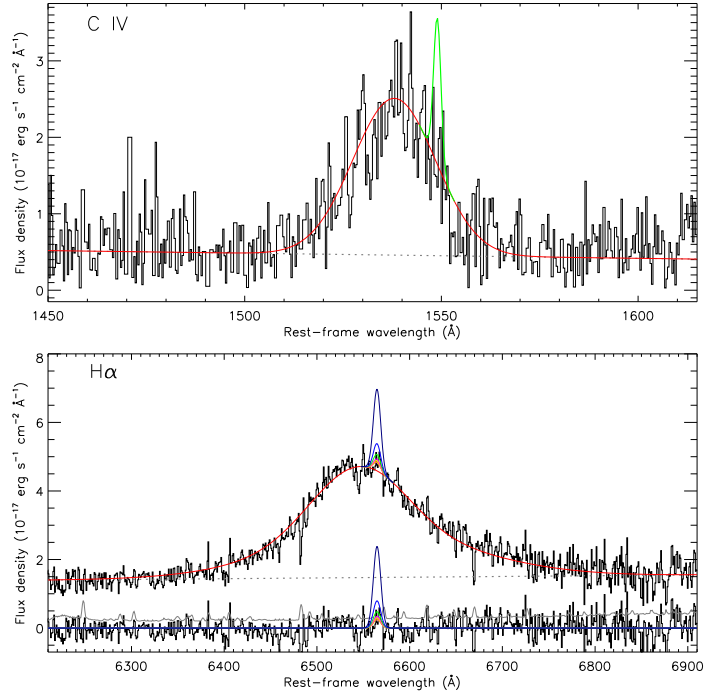}
\caption{Top panel: Theoretical \civ\ narrow line profile (green line) from the typical line ratio of Seyfert spectra.
Bottom panel: Theoretical \ha\ narrow line profiles (pink, orange, yellow magenta, green, blue, and navy lines)
with $f_{esc}(Ly\alpha)=90\%$, 80\%, 70\%, 50\%, 30\%, and 10\% in sequence,
overploted the observed spectrum and the best-fitting profile (black and red line).
The residual and observed error are also shown by black and gray lines for comparison.
}\label{f7}
\end{figure*}

\end{document}